\documentclass{spie}

\pdfoutput=1

\usepackage{amsmath,amsfonts,amssymb}
\usepackage{graphicx}
\usepackage[colorlinks=true, allcolors=blue]{hyperref}
\usepackage{placeins}
\usepackage{subfig}
\usepackage{graphicx}
\usepackage{caption}
\usepackage{placeins}
\usepackage{csquotes}
\usepackage{multirow}
\usepackage{tabu}

\title{BTFI2: a simple, light and compact Fabry-Perot instrument for the SOAR telescope}

\author[a]{Bruno C. Quint}
\author[b]{Brian Chinn} 
\author[c]{Claudia Mendes de Oliveira}
\author[d]{Philippe Amram} 
\author[c]{Denis Andrade}
\author[c,e]{William Schoenell}
\author[c]{Daniel Moser Faes}

\affil[a]{SOAR Telescope - La Serena, Chile}
\affil[b]{GEMINI Observatory - La Serena, Chile}
\affil[c]{Univ. de S\~ao Paulo - S\~ao~Paulo, Brazil}
\affil[d]{Aix-Marseille University, CNRS, LAM - Marseille, France}
\affil[e]{Univ. Federal do Rio Grande do Sul - Porto Alegre, Brazil}

\authorinfo{Further author information: (Send correspondence to B.C.Q.)\\B.C.Q.: E-mail: bquint@ctio.noao.edu, Telephone: 56 9 9935 5676}

\begin{document} 
\maketitle

\begin{abstract}
 
  We present the concept of a new Fabry-Perot instrument called BTFI-2, which is based on the design of another Brazilian instrument for the SOAR Telescope, the Brazilian Tunable Filter Imager (BTFI). BTFI-2 is designed to be mounted on the visitor port of the SOAR Adaptive Module (SAM) facility, on the SOAR telescope, at Cerro Pach\'on, Chile. This optical Fabry-Perot instrument will have a field of view of 3~x~3~arcmin, with 0.12~arcsec per pixel and spectral resolutions of 4500 and 12000, at H-alpha, dictated by the two ICOS Fabry-Perot devices available. The instrument will be unique for the study of centers of normal, interacting and active galaxies and the intergalactic medium, whenever spatial resolution over a large area is required. BTFI-2 will combine the best features of two previous instruments, SAM-FP and BTFI: it will use an Electron Multiplication detector for low and fast scanning, it will be built with the possibility of using a new Fabry-Perot etalon which provides a range of resolutions and it will be light enough to work attached to SAM, and hence the output data cubes will be GLAO-corrected.
  
\end{abstract}


\section{Introduction}
\label{sec:intro}  

	Numerous integral field spectrographs (IFS) have been deployed on 4 to 8 meter-class telescopes during the last few decades. These instruments are built to deliver 3D spectroscopy by sampling different regions of the sky at the same time using different techniques, principally: (1) lenslet arrays, e.g. TIGER/CFHT\cite{Bacon1995}, OASIS/CFHT and SAURON/WHT\cite{Bacon2001}, OSIRIS/Keck\cite{Larkin2006}. (2) mirror slicers, e.g. SPIFFI/VLT\cite{Eisenhauer2003}, NIFS/Gemini\cite{Hart2003}, GNIRS/Gemini\cite{Allington-Smith2006}, SWIFT\cite{Tecza2006}, SINFONI/VLT\cite{Eisenhauer2003b}, KMOS/VLT\cite{Content2006}, MUSE/VLT\cite{Laurent2010}. (3) fiber bundles, e.g. DensePak/KPNO\cite{Barden1988}, SILFID/CFHT\cite{Vanderriest1988}, INTEGRAL/WHT\cite{Arribas1998}, FLAME-GIRAFFE/VLT (Pasquini, 2003SPIE.4841.1682P) and more recently SAMI/AAO\cite{Croom2012}. These IFS have relatively wide spectral range and intermediate spectral resolution ($\sim$~1000~-~5000), but generally have small field-of-views (FoV).

	For larger FoV another class of instruments, the imaging spectrometers, are used. The most common types are the Fabry-Perot spectrometers (FPS) and Fourier Transform spectrometers (FTS). For a long time, large FoV scanning FPS have been developed in the optical, e.g. TAURUS\cite{Atherton1982}, CIGALE\cite{Boulesteix1984}, HIFI\cite{Bland1989}, MMTF\cite{Veilleux2010}, SAM-FP\cite{Oliveira2017}, and they have been used with great success, especially for projects which required small wavelength coverage and large FoV, of the order of a few arc-minutes squared. 

	This paper describes one such instrument, the Brazilian Tunable Filter Imager 2 (BTFI-2): a Fabry-Perot imaging spectrometer with high image quality provided by ground-layer adaptive optics (GLAO). This opens a whole new window for scientific exploration in areas such as the study of the physics of nearby galaxies, their kinematics, and their star formation history from pc to kpc scales by taking advantage of the large FoV (3~x~3 arcmin) and high spatial resolution, thanks to the SAM facility\cite{Tokovinin2016}. Galaxies consist of baryons distributed in multiple morphological and kinematic components (bulges, disks, bars, black holes) trapped into the deep potential well of dark matter halos. High resolution spatially resolved spectroscopy on large FoV allows gathering detailed spatial and spectral information to  measure numerous physical parameters such as (1) the interaction and feedback between the gas, the stars, the dust and the black holes on galactic (10-100 pc) scales and, (2) mass loss mechanisms in stars, HII regions, planetary nebulae and supernova on smaller (pc) scales for galactic structures. 

	BTFI-2 is a new instrument concept based on the Brazilian Tunable Filter Imager (BTFI, Ref. \citenum{Oliveira2013}), a FPS instrument designed for use with the SOAR Adaptive Module (SAM), at the 4.1~m SOAR Telescope, at Cerro Pach\'on, in Chile. BTFI is now officially decommissioned due to a number of reasons described further on in this paper. Instead, a much simpler instrument called SAM-FP has been used since 2016 to meet the needs of the SOAR community in obtaining spectroscopy with the SAM instrument. SAM-FP yields 3~$\times$~3~arcmin GLAO-corrected data cubes for a variety of studies, as mentioned in \autoref{sec:samfp.and.btfi}.  However, SAM-FP is equipped with a normal CCD, and not with a photon counting system like an EMCCD, given that it uses the regular SAM camera called SAMI, which limits the detector power at low flux of the instrument. Standard CCDs still exhibit readout noise of a few electrons per pixel which does not allow short exposures and furthermore fast scanning of elementary FP cycles for which sub-electron readout is requested (see \autoref{sec:performance} for details). The most important gain expected with BTFI-2 will be to get deeper data, thanks to the capability of scanning the data cube with shorter exposure times to averaging sky transmission fluctuations.
  
	Our plan for the new instrument BTFI-2 is to re-use part of the optical design and components of BTFI. BTFI-2 will then inherit the good features of both BTFI and SAM-FP: it will be a light instrument attached to the SAM's visitor's port (and therefore it will yield GLAO-corrected images), it will contemplate the use of a novel FPS that may enable a range of spectral resolutions, and it will use EMCCD (which will facilitate scanning and will allow for deeper data cubes to be taken). The unique feature of a FPS system attached to SAM is of course the GLAO-corrected images it delivers, over a FoV of 3$\times$3 arcmin. No other 4-meter class telescope has this capability.

    BTFI and SAM-FP, the two instruments which can be considered precursors of BTFI-2, are now briefly described in the \autoref{sec:samfp.and.btfi}. Then we we show the top level requirements for the instrument in \autoref{sec:top.level.requirements}. \autoref{sec:optical.design} describes the optical design. \autoref{sec:mechanical-design} describes the instrument's mechanical concept with a brief description of each element. Finally, \autoref{sec:conclusion} closes this work with some future steps and planning. 

\section{SAM-FP and BTFI, the precursors of BTFI2}
\label{sec:samfp.and.btfi}
   
    BTFI\cite{Oliveira2013} was an instrument designed and built at the University of S\`ao Paulo, Brazil to be used at the SOAR Telescope, either directly at the telescope's visitor port for seeing limited image quality or with SAM for enhanced image quality. It delivered high and low spectral resolution data cubes using Fabry-Perot interferometer and an imaging Bragg Tunable Filter (iBTF) in two different light paths. The iBTF was a new concept described at Ref. \cite{Blais-Ouellette2006} which used two identical volume-phase holographic gratings together acting as a tunable filter. For each channel BTFI had a EMCCD for fast readout and improved signal-to-noise ratio. The instrument was meant to be used with the THALES SESO Fabry-Perot, but those were not finished by the time that the instrument was commissioned. Instead, we had two ICOS FP that were used by that time with success, but with seeing limited images. 	
    
    Although we had several success runs, BTFI was decommissioned for several reasons. First, it did not comply with the size and weight requirements to be used at SAM's visitor port. Because of this, BTFI could only be used with seeing limited image quality. Second, even with successful tests in the lab, we never managed to have the desired quality data cubes using the iBTF. Finally, the EMCCDs started to present unexpected behavior in terms of sensitivity and noise. The EMCCDs were reviewed and are now fixed (see \autoref{ssec:emccds}), but we decided to focus on the use of the Fabry-Perot available in the collimated space of SAM in a new observation mode called SAM-FP.
    
	SAM-FP delivers data cubes with both spatial and spectral information by scanning FP while acquiring images using the SAM Imager (SAMI), which holds a classical CCD. SAM yields angular resolutions down to 0.5~arcsec over a FoV of of 3~x~3~arcmin in the optical, using Ground Layer Adaptive Optics (GLAO). The data cubes spectral resolution is determined by which of the two ICOS (Queensgate) Fabry-Perot's is available for use within SAM (see \autoref{ssec:fabry_perot}). 
    
  Since 2016A, SAM-FP has been producing high signal-to-noise GLAO-corrected data cubes of Galactic Planetary nebulae, HH objects, Giant HII regions and star-forming and active galaxies. Examples of data obtained with SAM-FP can be found in Ref. \citenum{Oliveira2017} and \citenum{Plana2017}. Although SAM-FP is producing excellent data, the large existing overhead times limit the use of the instrument for brighter objects, and the high noise levels of SAMI's detectors limits the use of the instrument for fainter targets (see \autoref{sec:performance}). 
     
  Considering these limitations, we want to replace the SAMI's CCD with the BTFI's EMCCD, which is much more appropriate for FP observations of faint sources like galaxies or faint nebulae since its readout noise is very low and its readout rate is fast, less than one second. With that, the EM detector will enable rapid scan in order to average the effects of the atmosphere and seeing during an acquisition. In addition, it will also allow going deeper, for a given exposure time, considering that this detector has better performance than a normal CCD for low-flux objects\cite{Daigle2009}. 
  
  This replacement requires a focal reducer because the physical dimensions of both detectors are different in terms of pixel size, number of pixels, and detector size. Considering that, the most reasonable option would be to re-use the BTFI's optical elements since they were designed for that purpose. Although we have the detector and the optical elements, BTFI-2 still needs an opto-mechanical concept that would follow the top level requirements. This concept is the focus of the work presented here.
  
\section{Top level requirements}
\label{sec:top.level.requirements}

	The first BTFI-2 top requirement is that it should be light enough to be installed on SAM. That means that the instrument itself cannot weigh more than $\sim$~100~kg (Eng.~Patr\'icio Schurter, private communication). This 100~kg limit does not consider computers and controllers because they can be connected to the instrument through cables instead of lying within the instrument. The mass of all optical components excluding the filter mechanism and any supports or stages for the components, is approximately 35~kg. 
    
	The volume requirements are provided by the SOAR staff in the form of a SolidWorks 3D model. All the work that we have done so far considered this available space envelope. Since computers and controllers were not considered in the instrument design, we still need to evaluate how and where we will put these components and how they will communicate with the instrument itself.
  
	Another top level constraint is that BTFI-2 will use the optical elements from its predecessor, BTFI. BTFI was designed and built to re-image (at F/7.1) the SOAR telescope native focal surface (at Nasmyth, F/16.5) and the focal surface resulting from the Adaptive Optics (AO) optical train. Its optical design provides images with 3.0 x 3.0 arcmin with 0.12 arcsec / pixel plate scale on a e2v CCD207 EMCCD 1696 x 1616 with 19 $\mu$m pixel. Since BTFI-2 will use the same optical elements and roughly the same optical design, these parameters should be kept the same. In particular, the focal reducer already used in BTFI will be needed here to match the field of BTFI-2 to that of SAM.
  
  The final spectral range is defined by the performance on the different elements. More specifically, this depends on the optical transmittance for SAM, the available Fabry-Perot's, BTFI's lenses and mirrors and the EMCCDs. BTFI was originally designed to operate between 3500 \AA\ and 7000 \AA\ but the SAM blue wavelength limit is 4000 \AA, with better performance towards the red. We do not currently have the transmission curves for the the two Queensgate etalons that we have at the moment.
  
  The BTFI-2 design had to minimize the number of fold mirrors to maximize optical throughput. We also had to consider access to the Fabry-Perot while on telescope for alignment purposes, a process that occurs at the beginning of each run. Easy access to the filter mechanism is also another requirement considering that narrow-band filters could be changed every night.   
  
\section{Optical design}
\label{sec:optical.design}

\begin{figure}[h!]
  \centering
  \includegraphics[width=0.95\textwidth]{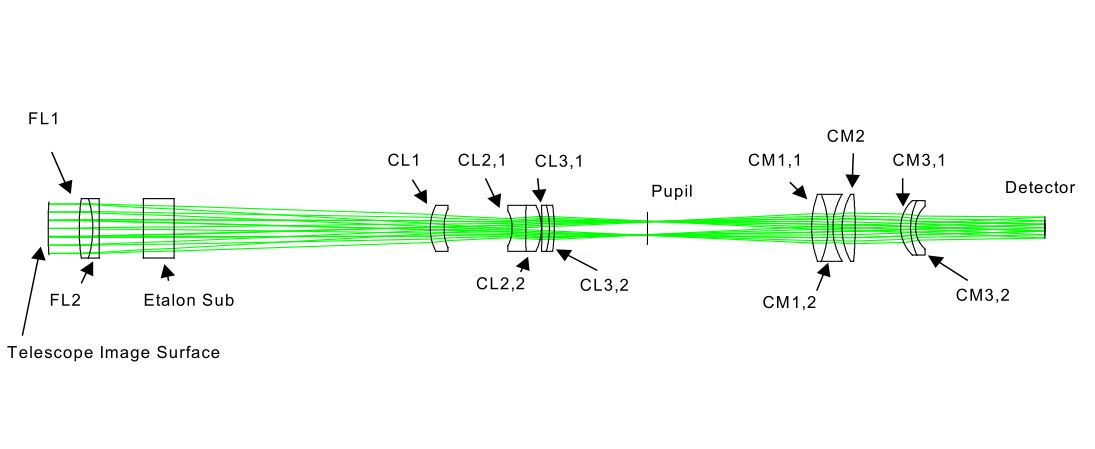}
  \centering
  \caption{The BTFI system layout. Details about this figure in the text.}
  \label{fig:btfi.layout}
  \end{figure}

  Conceptually, BTFI's design can be considered a simple imager with a focal reduction from F/16.5 to F/7.1. In practice, the instrument is much more complex. We show in \autoref{fig:btfi.layout} the optical designed by Opt. Eng. Damien Jones from the company Prime Optics without any folding mirror. In this figure, FL1 and FL2 are the field lenses, CL1 to CL3,1 are the collimator lenses and CM1,1 to CM3,2 are the camera lenses. The Etalon Sub shown in this figure is a glass plate that substitutes the low finesse Fabry-Perot that was supposed to lie at that position. 
    
  BTFI-2 uses the same optical design as the one shown in Fig. 1. Since BTFI is now retired, we plan to use its optical elements. Given that the base optical design is the same and we already have the optical elements in-hand, BTFI-2's overall cost goes down and the instrument is simpler to build.

  Although the basic optical design of BTFI-2 is the same, the optical path is different from BTFI due to the requirement to have a simpler, light and compact instrument. Light comes from SAM (or the telescope) with a F/16.25 ratio and meets the field lens (F$_{Field}$) after SAM's focal plane (P$_{obj}$), and a folding mirror (M$_1$) that makes the optical axis parallel to a bench, where the instrument optics are located. Then, we added another folding mirror (M$_2$) so the instrument can fit the available space envelope. After M$_2$, we have the collimator (L$_{Col}$), the pupil, the last folding mirror (M$_3$), the camera  lenses (L$_{Cam}$) and, finally, the detector. In the original BTFI, the low-finesse Fabry-Perot was substituted by a 50 mm thick dummy plate. Without it the total optical path is \~16 mm shorter. 
  
  \begin{figure}[h!]
  \centering
  \includegraphics[width=0.50\textwidth]{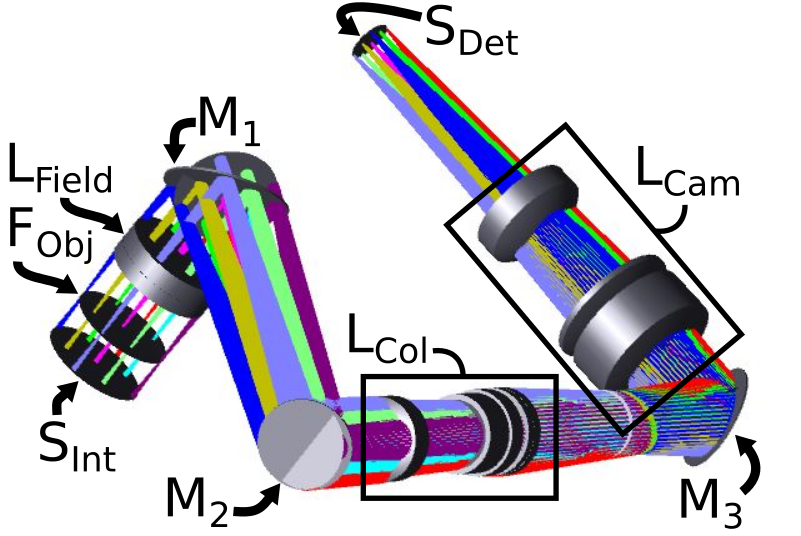}
  \centering
  \caption{The BTFI-2 system layout}
  \label{fig:layout-BTFI-2}
  \end{figure}

  We already have the conceptual design of the instrument bench with the mechanical supports for the lenses, the Fabry-Perot, the shutter and the filter wheel. The detector will be static and the focus will be set at the lab and minor adjustments will be done using shims. We plan to mount these elements at the CTIO's optical lab as a prototype and to help to review the design of the mechanical supports for the lenses. The idea is to have the procedure for optical alignment before we send any design for manufacturing.
  
  The conceptual mechanical design with the other elements is described below.

\section{Mechanical Design}  
\label{sec:mechanical-design}

  The mechanical design was driven primarily by the constraints described at \autoref{sec:top.level.requirements}. The base of the design is its structural subsystem, which includes an attachment ring that matches the SAM VI bolt pattern, and a bench that is fastened to this ring. All components are supported by the bench, and a lightweight sheet metal cover encloses the instrument components. Light enters the instrument with it's optical axis collinear to the central axis of the attachment ring. It then passes through the shutter and field lens, and is folded so that the optical axis is parallel to and 6.5cm above the surface of the bench.

  \begin{figure}[h!]
  \centering
  \includegraphics[width=0.60\textwidth]{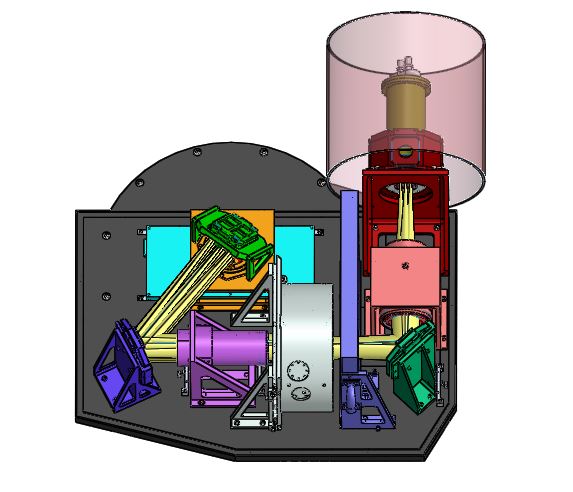}
  \caption{BTFI-2, with each of the ten components/subsystems highlighted in a different color. The first three subsystems, the shutter, field lens, and fold mirror A, are shown in light blue, orange, and bright green respectively}
  \label{fig:iso1}
  \end{figure}

  The light is then folded again before passing through the collimator. After the collimator, the light passes through the Fabry-Perot interferometer and interference filter\footnote{Please, check session \ref{sssec:filter-wheel} for further discussion on the filter wheel's position.} before being folded a third and final time. From the last fold, the light enters the camera before reaching the detector at the instrument focal plane.

  Access to the FP and the filter mechanism while the instrument is on the telescope were design drivers, and thus these components are accessible through a port in the instrument covers. The FP is mounted on slides, so that it can be removed from the inside of the covers for alignment access, and the filter mechanism can be removed entirely using this port as well. A more detailed description of this access is included in the next sub-section.
  
\subsection{Shutter}

  A 100 x 100 mm aperture Bonn Shutter is to be used near the visitor instrument focal plane of SAM, and is statically mounted to the bench. This is a rectangular aperture, high-precision, slit type shutter that comes with its own control electronics hardware and software, and is widely used within the astronomical community. Use of this shutter reduces development and testing costs/risks that would be associated with development of a shutter specifically for BTFI 2.0.
  
  Usually the shutter is located close to the detector but we decided to place it at the instrument's entrance to help protecting the instrument from dust, when not used. The shutter design is meant to keep the whole surface of the detector exposed at the same time even when placed close to the instrument focal plane.

\subsection{Field Lens}

  The field lens along with optical mount from BTFI are used. It is mounted statically to a support which is fastened to the bench, using pins to precisely define the position. The mount from BTFI includes adjustment along the optical axis via threads in the mount. The present alignment strategy does not require this degree of freedom, due to the relatively large axial tolerance for the field lens position. This feature will be included in BTFI-2 to keep the parts interchangeable between instruments.

\subsection{Fold Mirrors}

  The first fold mirror, fold mirror A, folds the beam so that the optical axis is parallel to the bench. All three fold mirrors are taken with their mounts from BTFI, which include tip-tilt adjustment. Folds B and C are identical, while fold A has a larger mirror surface area. The supports for these first two fold mirror/mount assemblies are static relative to the bench, and the fold C assembly includes axial adjustment.

\subsection{Collimator}

  The collimator and its optical mount from BTFI are used, with a new support structure that includes axial adjustment. Tip-tilt and de-centralization misalignment of the collimator are compensated for using the tip-tilt adjustment of the first two fold mirrors. Axial adjustment is accomplished using a push-pull design, with a M3 x 0.25 mm fine thread hex ball-nose adjuster, a M4 x 0.7 mm socket head cap screw to oppose the fine adjuster, a dovetail slide interface, and pre-load provided by spring-nose plungers.

\captionsetup{margin = 0pt}
\begin{figure}[ht]
\centering
\subfloat[]{
\label{fig:collimator}
\includegraphics[width=0.3\textwidth]{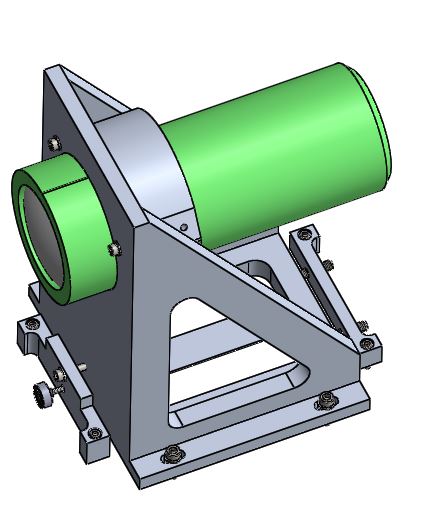}
} \qquad
\subfloat[]{
\label{fig:pushpull}
\includegraphics[width=0.3\textwidth]{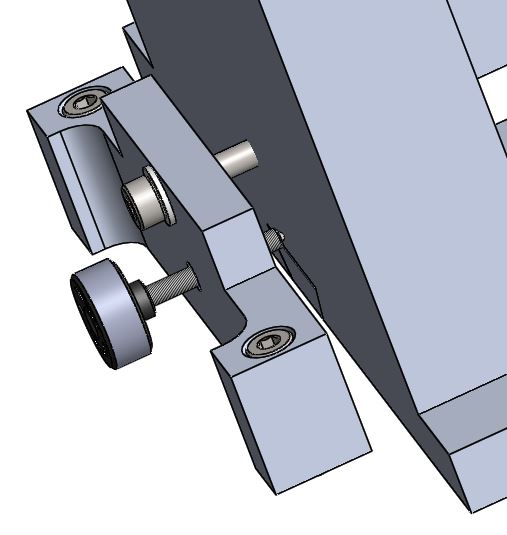}
}
\centering
\caption{(a) The collimator in its support. The support is on a dovetail slide, with axial position of the collimator actuated with the adjuster at the left in the image, and pre-load from spring nose plungers on the lower right in the image. (b) A detailed view of the adjustment actuator. The lower screw is a fine thread hex adjuster with knob that pushes the collimator in its support, and the upper screw is a socket head cap screw that pulls against the fine
thread adjuster. Once aligned, four screws in the support that secure the collimator to the bench are tightened to maintain alignment.}
\label{fig:collimatorsplit}
\end{figure}
\captionsetup{margin = 70pt}
\FloatBarrier
  
\subsection{Fabry-Perot}
\label{ssec:fabry_perot}

  Both BTFI and BTFI-2 are designed to work with the two new étalons that are being developed in France by Thales SESO. These new FP are required to have higher tunability (a few microns to hundreds of microns). We expect spectral  resolutions from 6000 to 25000. One of them is designed to have a higher Finesse so it will operates as classical FP.

  BTFI was designed to allocate both FP inside it, but BTFI-2 will have space for just one. This decision was made based on the weight and space constraints. The original idea was to have the two FPs working together so the first FP, with lower Finesse and close to BTFI's image plane, could work as a tunable order selector for the second FP, with higher Finesse and close to the pupil. In principle, this will not be allowed in BTFI-2 and we will have to perform observations using the conventional narrow-band filters as order selectors.
  
  One possibility not yet studied is to actually put the low order FP inside BTFI-2, to work as a tunable filter like described above and use one of the two ICOS Fabry-Perots that we have been using for the SAM-FP mode inside SAM. 

  SAM-FP is using the Fabry-Perot that we loaned from the University of Maryland and from the Australian Astronomical Observatory (AAO). The highest resolution FP in use in SAM-FP has a mean gap of 200~$\mu m$ and works at an interference order $p \simeq 609$ at H$\alpha$ 6562.78 \AA. A typical effective finesse $F \simeq 18$ and a resolution $R \simeq 11200$ (at H$\alpha$) are measured for this etalon. The second Queensgate Fabry-Perot device that can be used in SAM-FP has a gap of 44~$\mu m$ and a working interference order $ p\simeq 134$ at H$\alpha$ 6562.78 \AA. A typical effective finesse $F \simeq 32$ and a resolution $ R \simeq 4000$ (at H$\alpha$) are measured for this device. These parameters are put together on the \autoref{tab:fp-pars}. For early commissioning BTFI-2 could also use one of these FPs while the Thales SESO étalons are not finished.

  \begin{table}[!ht]
  \centering
  \caption{Nominal Parameters for the two Queensgate ET-70 Fabry-Perot}
  \begin{tabular}{l c c}
  \hline
  Gap Size                         & $44 \mu m$ & $200 \mu m$ \\
  Interference order for H$\alpha$ & 134        & 609         \\
  Resolution for H$\alpha$         & 4000		& 11200		  \\
  Free-Spectral-Range              & 48.95\AA   & 10.77\AA    \\
  Free-Spectral-Range [km/s]       & 2230 km/s  & 490 km/s    \\
  Finesse                          & 32         & 18          \\
  \hline
  \end{tabular}
  \label{tab:fp-pars}
  \end{table}  
  
  For BTFI-2, the FP is mounted on long-life, high-load rated slides, to allow manual access to the FP for alignment purposes while on the telescope without requiring removal of the FP from the instrument or the instrument from the telescope. A port in the covers can be removed for access, and captive screws are used to secure the FP in the "in beam" position to eliminate risk of dropping hardware into the instrument when accessing the FP when on the telescope. 

  The slides lock open with a lever release, to secure the FP during the alignment process. This process is manual, and the Fabry-Perot remains in the beam during normal operations. This makes it impossible to acquire images to check pointing while on sky, without the operator manually removing the FP from the beam. Initially, a motorized stage was considered in order to permit the acquisition of pointing images using BTFI-2's detector; however, this stage was eliminated due to mass limitations. Instead, pointing images will be obtained using SAMI, and pointing error corrected for using pre-calculated offsets between the two instruments. The process to switch between SAMI and BTFI-2 is simple, only requiring changing the orientation of the instrument selector mirror internal to SAM. This procedure does increase time requirements before image acquisition, but this additional time is insignificant when compared to the typical time required to scan with the FP.
  
\begin{figure}[h!]
\centering
\includegraphics[width=0.4\textwidth]{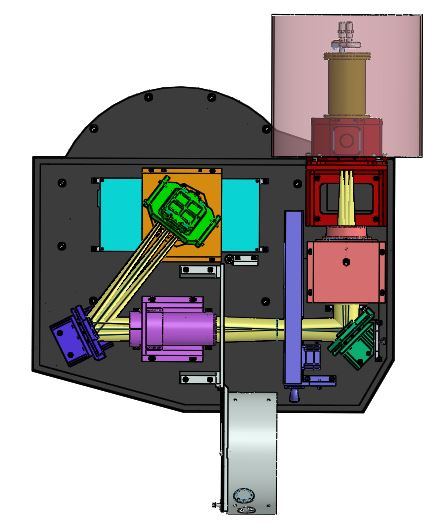}
\caption{The Fabry-Perot interferometer may be extended outside of the
  instrument covers for access for alignment purposes}
\label{fig:topextended}
\end{figure}

\captionsetup{margin = 0pt}
\begin{figure}[h!]
\centering
\subfloat[]{
  \label{fig:FPin}
  \includegraphics[width=0.3\textwidth]{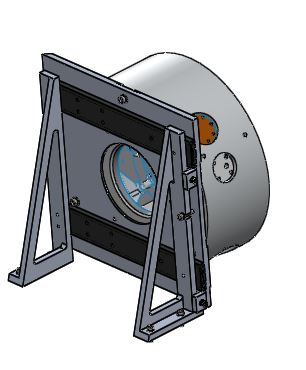}
  } \qquad
\subfloat[]{
  \label{fig:FPout}
  \includegraphics[width=0.4\textwidth]{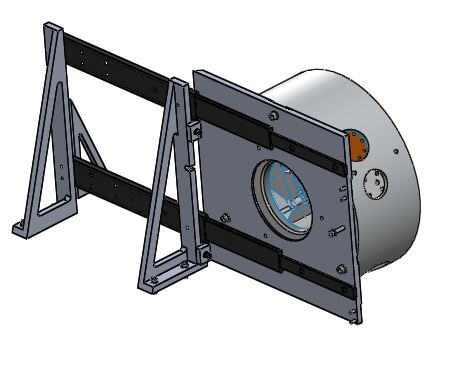}
  }
\caption{(a) The FP on its support in the "in beam" configuration. (b) The FP on its support in the "out of beam" orientation.}
\end{figure}
\captionsetup{margin = 70pt}
\FloatBarrier

\subsection{Filter Wheel}
\label{sssec:filter-wheel}

  The filter mechanism can be accessed through the same port in the covers as the FP, which allows the operator to remove the entire mechanism in order to change the filters. It is supported on a support rail, and linear ball bearings are used to couple the mechanism to the rail. The wheel has 6 apertures to allow for up 6 filters to be installed at a time. Due to the acquisition images being obtained with SAMI, it is not necessary to leave one of the apertures empty. Each filter is supported in a mechanical tilt assembly to allow for precision adjustment of the tilt of the filters. This allows for tuning of the central wavelengths of the interference filters. A fine adjuster with pitch of 0.20~mm is used to change the tilt, it is located at a distance from the axis of rotation of the tilt assembly such that 1~mm of linear extension corresponds to 1 degree of inclination of the the filters up to 5 degrees. A spring provides pre-load against the adjustment, and the assembly is secured in the desired position by two socket head cap screws.

\captionsetup{margin = 0pt}
\begin{figure}[h!]
      \centering
      \subfloat[]
        {
        \label{fig:angleasm1}
        \includegraphics[width=0.4\textwidth]{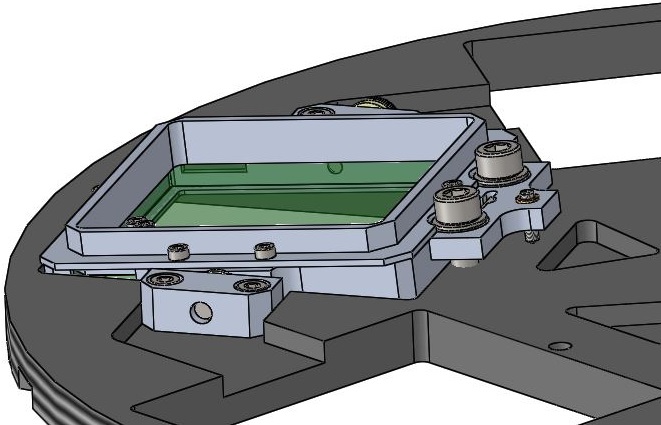}
        } \qquad
      \subfloat[]
        {
        \label{fig:angleasm2}
        \includegraphics[width=0.4\textwidth]{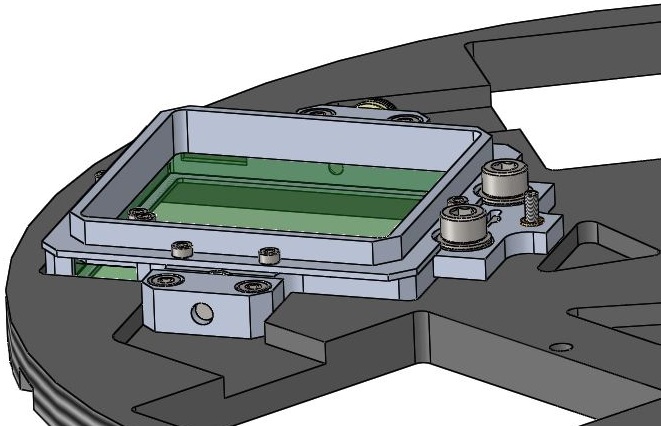}
        }
      \centering
      \caption{(a) The filter angle assembly inclined at an angle of 5 degrees. (b) The filter angle assembly in its nominal position, with 0 degrees inclination.}
\end{figure}
\captionsetup{margin = 70pt}
\FloatBarrier

  The wheel is driven by a pinion gear that engages to the outer diameter of the filter wheel, which has helical teeth machined on it's outer diameter. The pinion gear is driven by an Oriental Motor PKP564FN24AW high-torque stepper  motor. The wheel is supported at its central axis of rotation, and turns on bearings. It is held in a given position with a mechanical detent, which retains the wheel in the position defined by the user. A limit switch defines a homing position. This aspect of the design needs further development, as the gears are not yet fully defined.

\begin{figure}[h!]
    \centering
      \includegraphics[width=0.4\textwidth]{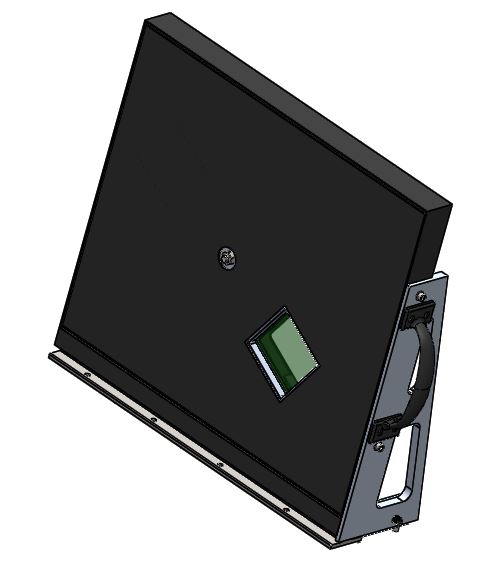}
      \centering
      \caption{The filter mechanism on its support rail. The angular piece on
the side is used to define the angle of the mechanism, and the handle provides
easy access for the operator. }
      \label{fig:filtermech}
\end{figure}

  The concept for filter wheel control is as follows. At the start of observations, the filter wheel will be commanded to go to the home position. Once this datum has been defined, commands changing the filter position will by determined by the number of steps between the initial and desired positions. Once in the desired position, current to the motor will be cut, and therefore the holding torque of the motor. The mechanical detent will then overcome parasitic torques in the system to retain the filter wheel in the desired position. Because this detent engagement is mechanical, filter positions should be precisely defined and lost steps will not affect position.

  The present design of the instrument has the interference filters located near the instrument pupil and after the FP, which is based on the design of BTFI. However, that will very likely cause multiple reflections between the reflective surface of the interference filters and the FP surfaces. To avoid that, the filter wheel should lie between the collimator and the FP. There may still some ghosting caused by reflections between the filter wheel and the collimator but they will be much weaker than the current design. 
    
\subsection{Camera}
  
\begin{figure}[h!]
\centering
\subfloat[]{
    \label{fig:camera}
    \includegraphics[width=0.3\textwidth]{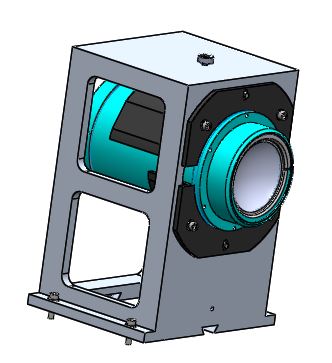}
    } \qquad
  \subfloat[]{
    \label{fig:detector}
    \includegraphics[width=0.3\textwidth]{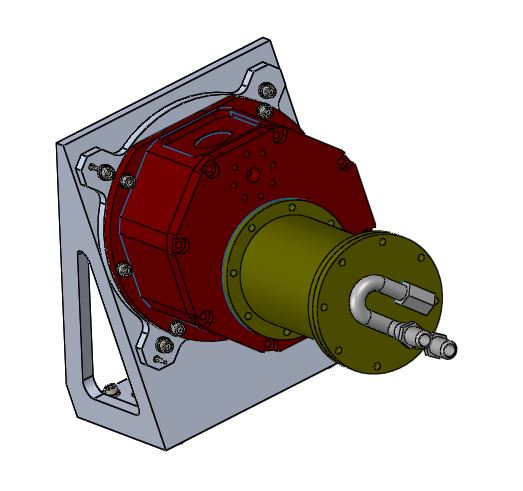}
    }
  \centering
  \caption{(a) The camera in its support. The knob at the top is used in the alignment process. (b) The detector dewar on its support.}
\end{figure}

  The camera has a linear degree of freedom along the optical axis, which is actuated using the same design adjustment for the collimator. Instead of adjusting the position of the detector dewar to focus, this adjustment is achieved with the camera, since the linear adjustment occurs in the collimated space. Another linear adjustment in the direction normal to the surface of the bench is used for decentralization correction (Figure \autoref{fig:camera}).
  
\subsection{EMCCDs}
\label{ssec:emccds}

  We have now two e2v CCD207 EMCCDs 1696 x 1616 px controlled by a CCD Controller for Counting Photons, or CCCP, each which were custom-built by a group based in the University of Montreal and is now commercialized by N\"uv\"u Cameras. These two detectors presented an unstable behavior in the past, which was one of the factors contributing to BTFI's retirement. The cameras' instability was related to terminal issues inside the camera, and this issue is now solved \cite{Andrade2016}. 

  The two detectors linearity up to 20 kADU for Camera 1 and up to 15 kADU for Camera 2 is guaranteed\cite{Andrade2016} for the classical mode. For the Amplified Mode, i.e. with eh EM gain turned on, the 65536 ADU digital limit imposed by the analogic-to-digital converter (ADC) is reached with much less light. The Quantum Efficiency in classic mode is  roughly constant and a bit higher than 90\% between 500 and 650 nm, and decreases almost linearly to $\sim$~35~\% at 900~nm and $\sim$~40~\% at 400~nm (\autoref{tab:qe}). 
  
  \begin{table}
  \begin{center}
  \caption{Quantum efficiency in classic mode for both cameras.}
  \begin{tabular}{c c c c}
  	\hline
  	Wavelength [\AA] & Data sheet [\%] & Camera 1 [\%] & Camera 2 [\%] \\ \hline
	350	& 20 & 12.92 & 16.62 \\ 
	400	& 52 & 46.84 & 49.49 \\ 
    500	& 90 & 84.46 & 86.54 \\
	650	& 90 & 86.45 & 87.78 \\
	900	& 37 & 31.73 & 32.11 \\ \hline
  \end{tabular}
  \label{tab:qe}
  \end{center}
  \end{table}
  
  For high-frame rate, the clock-induced charges (CIC) becomes dominant over the other sources of noise affecting the EMCCD (mainly dark noise)\cite{Daigle2008, Daigle2009}. Ref. \citenum{Andrade2016} makes a comparison between BTFI cameras and some other found in the literature and find out that the CIC values, shown in \autoref{tab:emccd_noises}, are very low. For example, Ref. \citenum{Bush2015} uses EMCCDs model CCD201-20 from E2V and obtained a CIC noise of 0.0447~$e^- \cdot px^{-1}$ with a Gain of 300. The commercial camera iXon3 860 from Andor company\footnote{\url{http://www.andor.com/scientific-cameras/ixon-emccd-camera-series/ixon3-860}} has a CIC noise of 0.05 $e^- \cdot px^{-1}$ for a Gain about the same level of our detectors. He also found some cases where the CIC noise was lower than ours, but those are equipment designed for specific goals and with smaller detectors (e.g., Daigle, 2009\cite{Daigle2009}). Although the two detectors still have to be tested on sky, the laboratory tests indicates that our detectors are now stable and that they could be used in their current condition on BTFI-2. 

  \begin{table}
  \begin{center}
  \caption{e2v CCD207 EMCCD Characteristics}
  \begin{tabular}{l c c}
  \\ \hline
                          & Camera 1                   & Camera 2 			\\ \hline
  Readout Noise           &1012 90.79 $e^- / px$           & 89.50 $e^- / px$ 	\\
  PreAmp Gain             & 3.86 $e^- / adu$ 		   & 4.71 $e^- / adu$ 	\\
  EM Gain                 & 1011 $e^- / e^-$           & 1005 $e^- / e^-$	\\
  CIC Noise               & 0.010 $e^- / px$           & 0.025 $e^- / px$ 	\\
  Effective Readout Noise & 0.1 $e^-$                  & 0.1 $e^-$ 			\\
  Dark noise              & 0.0039 $e^- \cdot px^{-1} \cdot s^{-1}$ 
  													   & 0.0045 $e^- \cdot px^{-1} \cdot s^{-1}$  
                                                       						\\ \hline 
  \multirow{2}{*}{Pixel size} & \multicolumn{2}{c}{19 $\mu m$} \\
  							  & \multicolumn{2}{c}{0.12 x 0.12 arcsec} \\
  \multirow{2}{*}{Detector Size} & \multicolumn{2}{c}{1696 x 1616} \\
  								 & \multicolumn{2}{c}{32.22 x 30.70 mm} \\
  Estimated average readout time & \multicolumn{2}{c}{247 ms for 10MHz} \\ \hline
  \end{tabular}
  \label{tab:emccd_noises}
  \end{center}
  \end{table}
   
	For use with BTFI-2, a new mechanical support was designed for the detectors (fig:detector). The detector dewar attaches to an interface piece, which in turn is fastened to the detector dewar support at four points. Three of these points include push-pull adjustment with fine adjusters to correct for tip-tilt misalignment. The three corners are oriented such that tip and tilt adjustments are isolated and there is not play between the adjustments. Spherical washers with socket head cap screws are used to ensure secure engagement after alignment.

\captionsetup{margin = 0pt}
\captionsetup{margin = 70pt}
\FloatBarrier

\section{Expected Performance}
\label{sec:performance}

  As mentioned before, one of the main motivations of having BTFI-2 is that we would like to have a better performance than SAM-FP. At the moment, SAM-FP is both limited by overhead time and a high readout noise. For observations of faint sources, SAMI's readout were higher than the sky noise. Data obtained in the last runs also showed a lot of noise artifacts like horizontal stripes that interfere severely on the data-cube analysis. 
   
    \autoref{tab:emccd_noises} shows that the readout noise on BTFI's detectors are much lower than SAMI's, with $1.45\  e^-/px$ readout noise, for 4$\times$4 binning in normal read speed with gain $2.1 e^- / adu$. The main noise source on EMCCDs is the CIC noise. For the detectors we have now, the CIC noise is $0.010\ e^-/px$ at a readout frequency 10 MHz (readout time $247\ ms$) and nominal EM Gain 1000. The CIC noise can make it less attractive for brighter sources but the required frame rate (few frames per second) would still keep it under the same levels as observing with a CCD but with the advantage of having multiple sweeps during the scanning process. 
    
	The average overhead time during SAM-FP observations is primarilty the sum of the readout time and the time spend to change the gap size of the étalon. The latest observations with SAMI-FP showed that a 4$\times$4 binning ($0.18\ \times\ 0.18\ arcsec$) is sufficient to properly sample the sky. We measured the readout time in this configuration using a set of BIAS images for different nights and we found an average value of $2.54\pm 0.10\ s$. The readout time for the BTFI's EMCCD can reach $500 ms$, but this alone does not guarantee that observations using them would be more efficient than with SAMI using a classical CCD.
  
  For the setup currently in use on SAM-FP, we would expect an overhead time of less than 500~ms, considering TCP/IP and serial communications between computer-controller-étalon when changing channels. We found that the actual average overhead time per channel during a FP scan changes with the exposure time. The longer the exposure time is, the smaller the overhead time will be. We believe that the source of this large overhead lies inside the software that controls the Fabry-Perot. Our experiments have shown that the lowest limit for the average overhead time is at exposure time of $\sim 140\ s$ which is limited by the readout time.   
  
  For the reasons above, we have used SAM-FP in \textit{slow scanning mode}, with typical exposures between 30 and 120 seconds per channel. This is a compromise between having the longest possible single-channel exposure time in order to have a sky-noise limited observation while not experiencing too much of a change in the observing conditions (seeing, airmass variations, passage of clouds, etc). The largest problem with the slow scanning mode is that any variation in flux or in the sky lines will be added to the cubes. As described on Ref. \citenum{Atherton1982}, one wants to scan as many times as possible with  rapid scan so the flux variation and the sky lines are averaged, have a better behavior and are easier to be removed from the data-cube. 
  
  \autoref{tab:overheads.single} shows the total overhead time with SAMI for one and for five sweeps while keeping the total exposure time per channel. It shows that observations with single scan with more than 60~s have a duty cycle loss of less than 12~\%, which is acceptable. The atmosphere and transparency effects are still important here. If we perform a five sweep scanning (scanning the cube five times) while keeping the total exposure time the same, the duty cycle reaches unpractical values. Considering the overhead of 500~ms as readout time of a EMCCD and 500~ms as the time between channels, the duty cycle loss for a total exposure of 30~s and five sweep is only 16~\%, 10 times more efficient than the same case with the CCD. 
  
  \begin{table}
  \begin{center}
  \caption{Duty cycle loss per channel on SAM-FP for different number of scans and different total exposure time per channel.}
  \begin{tabular}{c c c c} \hline
  Total time exposed per channel & Sweeps & Overhead per channel & Duty cycle loss \\ \hline
10 s & 1 & 9.43 s & 94.32 \% \\
30 s & 1 & 8.70 s & 28.98 \% \\
60 s & 1 & 7.59 s & 12.65 \% \\
90 s & 1 & 6.49 s & 7.21 \% \\
120 s & 1 & 5.38 s & 4.49 \% \\ \hline
10 s & 5 & 48.63 s & 486.30 \% \\
30 s & 5 & 47.89 s & 159.65 \% \\
60 s & 5 & 46.79 s & 77.98 \% \\
90 s & 5 & 45.69 s & 50.76 \% \\
120 s & 5 & 44.58 s & 37.15 \% \\
  \hline
  \end{tabular}	
  \label{tab:overheads.single}
  \end{center}
  \end{table}
  
    Finally, we expect an optical performance similar to BTFI. Ref.~\citenum{Quint2015} measures BTFI's imaging quality using a 50~$\mu$m pinhole at the instrument focal plane. The expected spot size produced at the detector was 1.12~pixel diameter while it was measured had 2.45~pixels (0.30~arcsec) radius for Camera~1 and 2.14~pixel (0.26~arcsec) radius for Camera~2. This is not an optimal value since SAM can provide image with up to 0.4~arcsec during nights with optical conditions. Most frequently, the delivered fwhm is about 0.5~-~0.6~arcsec under good conditions so this is not a killing issue.

\section{Conclusion}
\label{sec:conclusion}

	This document shows that BTFI-2 is a feasible project considering the optical elements, optical and mechanical concept, and detectors are all available. Accordingly to the expected performance described, BTFI-2 will expand the limits of Fabry-Perot science that can be done at SOAR with SAM, both by increasing the sensitivity and lowering down the duty cycle loss.     
        
    We plan to improve the FP control system to reach the 500~ms overhead time using the two loaned etalons. The THALES SESO FP must also fit this requirement. A control system will be developed using the Chimera Observatory Control System\cite{Silva2017}\footnote{\url{https://github.com/astroufsc/chimera}}, a Python Framework developed to manage observatories and astronomical instruments.
    
    Finally, the BTFI-2 proposal is aligned to the future of instrumentation in SOAR. An upgrade to SAM is under way to improve the performance of GLAO correction at shorter wavelengths in the visible\cite{Faes2018}. This enhances BTFI-2 to make line-ratio studies and reddening measurements with good angular accuracy at (almost) all visible range over the large FoV provided by SOAR.
    
\acknowledgments 
 
BQ acknowledges funding from CNPq fellowship 205459/2014-5. DMF acknowledges support from FAPESP 2016/16844-1. CMdO and PA are thankful to the SPRINT program of FAPESP (2017/50277-0) and CNRS, which has made SAM-FP tests and observing runs possible. CMdO also acknowledges funding from FAPESP grants 2014/07684-5, 2016/17119-9 and CNPq grant 312333/2014-5.

We thank our colleagues from SOAR Telescope who provided insight and expertise that greatly assisted the research. We also thank Dr. Andrei Tokovinin for assistance with the comparisons between the performance of SAMI and the expected performance on the EMCCDs.


\begin{thebibliography}{10}

\bibitem{Bacon1995}
{Bacon}, R., {Adam}, G., {Baranne}, A., {Courtes}, G., {Dubet}, D., {Dubois},
  J.~P., {Emsellem}, E., {Ferruit}, P., {Georgelin}, Y., {Monnet}, G.,
  {Pecontal}, E., {Rousset}, A., and {Say}, F., ``{3D spectrography at high
  spatial resolution. I. Concept and realization of the integral field
  spectrograph TIGER.},'' {\em Astronomy and Astrophysics Supplement
  Series}~{\bf 113},  347 (Oct. 1995).

\bibitem{Bacon2001}
{Bacon}, R., {Copin}, Y., {Monnet}, G., {Miller}, B.~W., {Allington-Smith},
  J.~R., {Bureau}, M., {Carollo}, C.~M., {Davies}, R.~L., {Emsellem}, E.,
  {Kuntschner}, H., {Peletier}, R.~F., {Verolme}, E.~K., and {de Zeeuw}, P.~T.,
  ``{The SAURON project - I. The panoramic integral-field spectrograph},'' {\em
  Monthly Notices of the Royal Astronomical Society}~{\bf 326},  23--35 (Sept.
  2001).

\bibitem{Larkin2006}
{Larkin}, J., {Barczys}, M., {Krabbe}, A., {Adkins}, S., {Aliado}, T., {Amico},
  P., {Brims}, G., {Campbell}, R., {Canfield}, J., {Gasaway}, T., {Honey}, A.,
  {Iserlohe}, C., {Johnson}, C., {Kress}, E., {LaFreniere}, D., {Lyke}, J.,
  {Magnone}, K., {Magnone}, N., {McElwain}, M., {Moon}, J., {Quirrenbach}, A.,
  {Skulason}, G., {Song}, I., {Spencer}, M., {Weiss}, J., and {Wright}, S.,
  ``{OSIRIS: a diffraction limited integral field spectrograph for Keck},'' in
  [{\em Ground-based and Airborne Instrumentation for Astronomy. Edited by
  McLean, Ian S.; Iye, Masanori. Proceedings of the SPIE, Volume 6269, id.
  62691A (2006).}{\nolinebreak\hspace{0.1em}]},   {\bf 6269} (June 2006).

\bibitem{Eisenhauer2003}
{Eisenhauer}, F., {Tecza}, M., {Thatte}, N., {Genzel}, R., {Abuter}, R.,
  {Iserlohe}, C., {Schreiber}, J., {Huber}, S., {Roehrle}, C., {Horrobin}, M.,
  {Schegerer}, A., {Baker}, A.~J., {Bender}, R., {Davies}, R., {Lehnert}, M.,
  {Lutz}, D., {Nesvadba}, N., {Ott}, T., {Seitz}, S., {Schoedel}, R.,
  {Tacconi}, L.~J., {Bonnet}, H., {Castillo}, R., {Conzelmann}, R.,
  {Donaldson}, R., {Finger}, G., {Gillet}, G., {Hubin}, N., {Kissler- Patig},
  M., {Lizon}, J.~L., {Monnet}, G., and {Stroebele}, S., ``{The Universe in 3D:
  First Observations with SPIFFI, the Infrared Integral Field Spectrometer for
  the VLT},'' {\em The Messenger}~{\bf 113},  17--25 (Sept. 2003).

\bibitem{Hart2003}
{Hart}, J., {McGregor}, P.~J., and {Bloxham}, G.~J., ``{NIFS concentric
  integral field unit},'' in [{\em Instrument Design and Performance for
  Optical/Infrared Ground-based Telescopes. Edited by Iye, Masanori; Moorwood,
  Alan F. M. Proceedings of the SPIE, Volume 4841, pp. 319-329
  (2003).}{\nolinebreak\hspace{0.1em}]},   {\bf 4841},  319--329 (Mar. 2003).

\bibitem{Allington-Smith2006}
{Allington-Smith}, J.~R., {Content}, R., {Dubbeldam}, C.~M., {Robertson},
  D.~J., and {Preuss}, W., ``{New techniques for integral field spectroscopy -
  I. Design, construction and testing of the GNIRS IFU},'' {\em Monthly Notices
  of the Royal Astronomical Society}~{\bf 371},  380--394 (Sept. 2006).

\bibitem{Tecza2006}
{Tecza}, M., {Thatte}, N., {Clarke}, F., {Goodsall}, T., {Freeman}, D., and
  {Salaun}, Y., ``{SWIFT image slicer: large format, compact, low scatter image
  slicing},'' in [{\em Optomechanical Technologies for Astronomy. Edited by
  Atad-Ettedgui, Eli; Antebi, Joseph; Lemke, Dietrich. Proceedings of the SPIE,
  Volume 6273, id. 62732L (2006).}{\nolinebreak\hspace{0.1em}]},   {\bf 6273}
  (June 2006).

\bibitem{Eisenhauer2003b}
{Eisenhauer}, F., {Abuter}, R., {Bickert}, K., {Biancat-Marchet}, F., {Bonnet},
  H., {Brynnel}, J., {Conzelmann}, R.~D., {Delabre}, B., {Donaldson}, R.,
  {Farinato}, J., {Fedrigo}, E., {Genzel}, R., {Hubin}, N.~N., {Iserlohe}, C.,
  {Kasper}, M.~E., {Kissler-Patig}, M., {Monnet}, G.~J., {Roehrle}, C.,
  {Schreiber}, J., {Stroebele}, S., {Tecza}, M., {Thatte}, N.~A., and {Weisz},
  H., ``{SINFONI - Integral field spectroscopy at 50 milli-arcsecond resolution
  with the ESO VLT},'' in [{\em Instrument Design and Performance for
  Optical/Infrared Ground-based Telescopes. Edited by Iye, Masanori; Moorwood,
  Alan F. M. Proceedings of the SPIE, Volume 4841, pp. 1548-1561
  (2003).}{\nolinebreak\hspace{0.1em}]},   {\bf 4841},  1548--1561 (Mar. 2003).

\bibitem{Content2006}
{Content}, R., ``{Optical design of the KMOS slicer system},'' in [{\em
  Ground-based and Airborne Instrumentation for Astronomy. Edited by McLean,
  Ian S.; Iye, Masanori. Proceedings of the SPIE, Volume 6269, id. 62693S
  (2006).}{\nolinebreak\hspace{0.1em}]},   {\bf 6269} (June 2006).

\bibitem{Laurent2010}
{Laurent}, F., {Adjali}, L., {Arns}, J., {Bacon}, R., {Boudon}, D., {Caillier},
  P., {Daguis{\'e}}, E., {Delabre}, B., {Dubois}, J.-P., {Godefroy}, P.,
  {Jarno}, A., {Jorden}, P., {Kosmalski}, J., {Lap{\`e}re}, V., {Lizon}, J.-L.,
  {Loupias}, M., {Pecontal}, A., {Reiss}, R., {Remillieux}, A., {Renault}, E.,
  {Rupprecht}, G., and {Salaun}, Y., ``{MUSE integral field unit: test results
  on the first out of 24},'' in [{\em Proceedings of the SPIE, Volume 7739, id.
  77394M (2010).}{\nolinebreak\hspace{0.1em}]},   {\bf 7739} (July 2010).

\bibitem{Barden1988}
{Barden}, S.~C. and {Wade}, R.~A., ``{DensePak and spectral imaging with fiber
  optics.},'' in [{\em IN: Fiber optics in astronomy; Proceedings of the
  Conference, Tucson, AZ, Apr. 11-14, 1988 (A90-20901 07-35). San Francisco,
  CA, Astronomical Society of the Pacific, 1988, p.
  113-124.}{\nolinebreak\hspace{0.1em}]},   {\bf 3},  113--124 (Jan. 1988).

\bibitem{Vanderriest1988}
{Vanderriest}, C. and {Lemonnier}, J.~P., ``{Silfid - a Versatile Fiber-Optics
  Spectrograph for Faint Objects},'' in [{\em Instrumentation for Ground-Based
  Optical Astronomy, Present and Future. The Ninth Santa Cruz Summer Workshop
  in Astronomy and Astrophysics, July 13- 24, 1987, Lick Observatory. Editor,
  L.B. Robinson; Publisher, Springer-Verlag, New York, NY, 1988. LC QB856 .S26
  1987. ISBN 0-387-96730-3. P.304, 1988}{\nolinebreak\hspace{0.1em}]},   304
  (Jan. 1988).

\bibitem{Arribas1998}
{Arribas}, S., {Carter}, D., {Cavaller}, L., {del Burgo}, C., {Edwards}, R.,
  {Fuentes}, F.~J., {Garcia}, A.~A., {Herreros}, J.~M., {Jones}, L.~R.,
  {Mediavilla}, E., {Pi}, M., {Pollacco}, D., {Rasilla}, J.~L., {Rees}, P.~C.,
  and {Sosa}, N.~A., ``{INTEGRAL: a matrix optical fiber system for WYFFOS},''
  in [{\em Proc. SPIE Vol. 3355, p. 821-827, Optical Astronomical
  Instrumentation, Sandro D'Odorico; Ed.}{\nolinebreak\hspace{0.1em}]},   {\bf
  3355},  821--827 (July 1998).

\bibitem{Croom2012}
{Croom}, S.~M., {Lawrence}, J.~S., {Bland-Hawthorn}, J., {Bryant}, J.~J.,
  {Fogarty}, L., {Richards}, S., {Goodwin}, M., {Farrell}, T., {Miziarski}, S.,
  {Heald}, R., {Jones}, D.~H., {Lee}, S., {Colless}, M., {Brough}, S.,
  {Hopkins}, A.~M., {Bauer}, A.~E., {Birchall}, M.~N., {Ellis}, S., {Horton},
  A., {Leon-Saval}, S., {Lewis}, G., {L{\'o}pez-S{\'a}nchez}, {\'A}.~R., {Min},
  S.-S., {Trinh}, C., and {Trowland}, H., ``{The Sydney-AAO Multi-object
  Integral field spectrograph},'' {\em Monthly Notices of the Royal
  Astronomical Society}~{\bf 421},  872--893 (Mar. 2012).

\bibitem{Atherton1982}
Atherton, P.~D., Taylor, K., Pike, C.~D., Harmer, C. F.~W., Parker, N.~W., and
  Hook, R.~N., ``{TAURUS: a wide-field imaging Fabry-Perot spectrometer for
  astronomy},'' {\em Monthly Notices of the Royal Astronomical Society}~{\bf
  201},  661 -- 696 (1982).

\bibitem{Boulesteix1984}
{Boulesteix}, J., {Georgelin}, Y., {Marcelin}, M., and {Monnet}, G., ``{First
  results from CIGALE scanning Perot-Fabry interferometer.},'' in [{\em IN:
  Instrumentation in astronomy V; Proceedings of the Fifth Meeting, London,
  England, September 7-9, 1983 (A85-25360 10-89). Bellingham, WA, SPIE - The
  International Society for Optical Engineering, 1984, p. 37-41. Research
  supported by the Institut National d'Astronomie et de
  Geophysique.}{\nolinebreak\hspace{0.1em}]},   {\bf 445},  37--41 (Jan. 1984).

\bibitem{Bland1989}
{Bland}, J. and {Tully}, R.~B., ``{The Hawaii imaging Fabry-Perot
  interferometer (HIFI)},'' {\em The Astronomical Journal}~{\bf 98},  723--735
  (Aug. 1989).

\bibitem{Veilleux2010}
{Veilleux}, S., {Weiner}, B.~J., {Rupke}, D.~S.~N., {McDonald}, M., {Birk}, C.,
  {Bland-Hawthorn}, J., {Dressler}, A., {Hare}, T., {Osip}, D.,
  {Pietraszewski}, C., and {Vogel}, S.~N., ``{MMTF: The Maryland-Magellan
  Tunable Filter},'' {\em The Astronomical Journal}~{\bf 139},  145--157 (Jan.
  2010).

\bibitem{Oliveira2017}
{Mendes de Oliveira}, C., {Amram}, P., {Quint}, B.~C., {Torres-Flores}, S.,
  {Barb{\'a}}, R., and {Andrade}, D., ``{First results from SAM-FP: Fabry-Perot
  observations with ground-layer adaptive optics - the structure and kinematics
  of the core of 30 Doradus},'' {\em Monthly Notices of the Royal Astronomical
  Society}~{\bf 469},  3424--3443 (Aug. 2017).

\bibitem{Tokovinin2016}
{Tokovinin}, A., {Cantarutti}, R., {Tighe}, R., {Schurter}, P., {Martinez}, M.,
  {Thomas}, S., and {van der Bliek}, N., ``{SOAR Adaptive Module (SAM): Seeing
  Improvement with a UV Laser},'' {\em PASP}~{\bf 128},  125003 (Dec. 2016).

\bibitem{Oliveira2013}
{Mendes de Oliveira}, C., {Taylor}, K., {Quint}, B., {Andrade}, D., {Ferrari},
  F., {Laporte}, R., {Ramos}, G. d.~A., {Dani Guzman}, C., {Cavalcanti}, L.,
  {de Calasans}, A., {Ramirez Fernandez}, J., {Gutierrez Casta{\~n}eda}, E.~C.,
  {Jones}, D., {Fontes}, F.~L., {Molina}, A.~M., {Fialho}, F., {Plana}, H.,
  {Jablonski}, F.~J., {Reitano}, L., {Daigle}, O., {Scarano}, S., {Amram}, P.,
  {Balard}, P., {Gach}, J.-L., and {Carignan}, C., ``{The Brazilian Tunable
  Filter Imager for the SOAR Telescope},'' {\em Publications of the
  Astronomical Society of the Pacific}~{\bf 125},  396 (Apr. 2013).

\bibitem{Blais-Ouellette2006}
{Blais-Ouellette}, S., {Daigle}, O., and {Taylor}, K., ``{The imaging Bragg
  tunable filter: a new path to integral field spectroscopy and narrow band
  imaging},'' in [{\em Ground-based and Airborne Instrumentation for Astronomy.
  Edited by McLean, Ian S.; Iye, Masanori. Proceedings of the SPIE, Volume
  6269, id. 62695H (2006).}{\nolinebreak\hspace{0.1em}]},   {\bf 6269} (June
  2006).

\bibitem{Plana2017}
{Plana}, H., {Rampazzo}, R., {Mazzei}, P., {Marino}, A., {Amram}, P., and
  {Ribeiro}, A.~L.~B., ``{The NGC 454 system: anatomy of a mixed ongoing
  merger},'' {\em Monthly Notices of the Royal Astronomical Society}~{\bf 472},
   3074--3092 (Dec. 2017).

\bibitem{Daigle2009}
{Daigle}, O., {Carignan}, C., {Gach}, J.-L., {Guillaume}, C., {Lessard}, S.,
  {Fortin}, C.-A., and {Blais-Ouellette}, S., ``{Extreme Faint Flux Imaging
  with an EMCCD},'' {\em Publications of the Astronomical Society of the
  Pacific}~{\bf 121},  866 (Aug. 2009).

\bibitem{Andrade2016}
Andrade, D. F. D.~E., {\em {Otimiza{\c{c}}{\~{a}}o das c{\^{a}}meras
  astron{\^{o}}micas do instrumento Brazilian Tunable Filter Imager}}, tese,
  Escola Polit{\'{e}}cnica - Universidade de S{\~{a}}o Paulo (2016).

\bibitem{Daigle2008}
{Daigle}, O., {Gach}, J.-L., {Guillaume}, C., {Lessard}, S., {Carignan}, C.,
  and {Blais-Ouellette}, S., ``{CCCP: a CCD controller for counting photons},''
  in [{\em Ground-based and Airborne Instrumentation for Astronomy II. Edited
  by McLean, Ian S.; Casali, Mark M. Proceedings of the SPIE, Volume 7014,
  article id. 70146L, 10 pp. (2008).}{\nolinebreak\hspace{0.1em}]},   {\bf
  7014} (July 2008).

\bibitem{Bush2015}
Bush, N., Hall, D., Holland, A., Burgon, R., Murray, N., Gow, J., Soman, M.,
  Jordan, D., Demers, R., Harding, L., Hoenk, M., Michaels, D., Nemati, B., and
  Peddada, P., ``{The impact of radiation damage on photon counting with an
  EMCCD for the WFIRST-AFTA coronagraph},'' {\em Proceedings of SPIE - The
  International Society for Optical Engineering}~{\bf 9605} (2015).

\bibitem{Quint2015}
Quint, B.~C., {\em {Estudo e caracteriza{\c{c}}{\~{a}}o do instrumento
  Brazilian Tunable Filter Imager}}, tese de doutorado, Universidade de
  S{\~{a}}o Paulo (2015).

\bibitem{Silva2017}
Silva, P.~H., Schoenell, W., Ribeiro, T., and Kanaan, A., ``astroufsc/chimera:
  Version 1.0,'' (May 2017).

\bibitem{Faes2018}
{Faes}, D.~M., {Tokovinin}, A., {Vieira}, T., {Mello}, A., {Domingues}, M.,
  {Andrade}, D., {Souza}, A., {Quint}, B.~C., {Santos}, J.~B., and {Almeida},
  M. C.~P., ``{SAMplus: adaptive optics in optical wavelengths at SOAR},'' in
  [{\em Ground-based and Airborne Instrumentation for Astronomy VII, Proc.
  SPIE}{\nolinebreak\hspace{0.1em}]},   {\bf this conference} (2018).

\end{thebibliography}

\bibliographystyle{spiebib} 

\end{document}